\documentclass[encvountsame]{llncs}

\usepackage{amsmath}

\begin{document}
    \title{Security Weakness of Flexible Group Key Exchange with On-Demand Computation of Subgroup Keys}
    \author {Qingfeng Cheng,   Chuangui Ma}
    \institute {Zhengzhou Information Science and Technology
      Institute, \\
    Zhengzhou, P.\,R.\,China\\
     \email{qingfengc2008@sina.com}
}


  \maketitle
    \begin{abstract}
In AFRICACRYPT 2010, Abdalla et al. first proposed a slight
modification to the computations steps of the BD protocol, called
mBD+P. Then they extended mBD+P protocol into mBD+S protocol. In
this paper, we show that both of mBD+P and mBD+S protocols are
vulnerable to malicious insiders attack. Further, we propose a
simple countermeasure against this attack.

    \end{abstract}


  \section{Introduction}
    \label{Introduction}
 Group key exchange (GKE) enables three or more parties to agree upon
a common secret session key in the open network for secure group
communication. However, GKE protocols is currently less well
understood than the case of two-party key exchange protocols. Many
security attributes have so far been ignored for the case of GKE
protocols.


In 2009, Manulis proposed flexible GKE protocols \cite{Manulis}
utilizing the well-known parallel Diffie-Hellman key exchange
(PDHKE) technique in which each party uses the same exponent for the
computation of peer-to-peer (p2p) keys with its peers. Further,
Manulis investigated possible optimizations of these protocols
allowing parties to re-use their exponents to compute both group and
p2p keys, and showed that not all such GKE protocols could be
optimized, which included the original Burmester-Desmedt (BD) GKE
protocol \cite{BD}.

Recently, Abdalla et al. used the more generalized and flexible
approach than Manulis's scheme to propose two GKE protocols: mBD+P
and mBD+S \cite{FGKE}, which are based on the well-studied BD GKE
protocol. The mBD+P protocol is modified for obtaining the secure
merge of BD and PDHKE. The mBD+S protocol as the extension of the
mBD+P protocol gets the ability to compute an independent session
key for any possible subgroup of the initial GKE users. In addition,
the authentication procedure in their protocols is similar to the
general authentication technique from \cite{Katz} and both of mBD+P
and mBD+S protocols are proven the security in the random oracle
model. In this paper, we will show that their protocols are
vulnerable to malicious insider attack. Under our attack, malicious
insiders can disrupt establishment of a common group session key
among all group members. Furthermore, we improve their protocols and
use key confirmation technique to overcome this secure flaw.

The rest of this paper is organized as follows. In Section 2, we
briefly review Abdalla et al.'s protocols. In Section 3, we show
that their protocols can't resist malicious insiders attack. In
Section 4, we propose our improvement to repair this secure flaw.
Finally, the conclusions will be given in Section 5.


\section{Review of mBD+P and mBD+S Protocols}\label{REV}

In this section, we briefly review mBD+P and mBD+S protocols
proposed by Abdalla et al. in 2010. In Table 1, we list the
abbreviations and notations used in mBD+P and mBD+S protocols. For
more details, we refer to \cite{FGKE}.

\begin{table}[h]
 \centering
  \caption{The notations}

\begin{tabular}{l l}
  \hline
  Notations & \ \ \ Description \\
  \hline
  $q$ &\ \ \  A large prime\\
  $\tau$ &\ \ \  Security parameter\\
  $G$&\ \ \ A cyclic additive group of order $q$\\
  $H_g,H_p,H_s$&\ \ \ Random oracles from $\{0,1\}^{*}$ to $\{0,1\}^{\tau}$\\
  $H$&\ \ \ Random oracle from $G$ to $\{0,1\}^{\tau}$\\
  $n$&\ \ \ The number of users\\
  $ U_1,U_2,...,U_{n-1},U_n$&\ \ \ Users \\
  $Sign$&\ \ \ A digital signature scheme\\
  $sk_i$&\ \ \ Signature private key\\
  $pk_i$&\ \ \ Verification public key\\

 \hline
\end{tabular}
 \end{table}

\subsection{mBD+P Protocol}

In this subsection, we briefly review the mBD+P protocol, which
includes two stages: group stage and p2p stage. On the correctness
of key computation and the security analysis of the mBD+P protocol
refer to \cite{FGKE}.

\subsubsection{Group Stage}


Let the group users be defined by \textbf{pid}=$(U_1,...,U_n)$. In
the following description we assume that user indices form a cycle
such that $U_0=U_n$ and $U_{n+1}=U_1$.
\begin{flushleft}

 \textbf{[Round 1]}. Each $U_i$ computes $y_i=g^{x_i}$ for some random
$x_i \in_R Z_q$ and broadcasts $(U_i,y_i)$.

\textbf{[Round 2]}. Each $U_i$ proceeds as follows:

\begin{itemize}
\item lets $sid_i=(U_1|y_1,...,U_n|y_n)$,

\item computes $k_{i-1,i}^{\prime}=y_{i-1}^{x_i}$ and
$k_{i,i+1}^{\prime}=y_{i+1}^{x_i}$,

\item $z_{i-1,i}^{\prime}=H(k_{i-1,i}^{\prime},sid_i)$ and
$z_{i,i+1}^{\prime}=H(k_{i,i+1}^{\prime},sid_i)$,

\item $z_i=z_{i-1,i}^{\prime}\oplus z_{i,i+1}^{\prime}$,

\item $\sigma_i=Sign(sk_i,(U_i,z_i,sid_i))$,

\item broadcasts $(U_i,z_i,\sigma_i)$.

\end{itemize}

\textbf{[Group Key Computation]}. Each $U_i$ checks whether
$z_1\oplus ... \oplus z_n=0$ and whether all received signatures
$\sigma_j$ are valid and aborts if any of these checks fails.
Otherwise, $U_i$ proceeds as follows:

\begin{itemize}
\item iteratively for each $j=i,...,i+n-1$, computes $z_{j,j+1}^{\prime}=z_{j-1,j}^{\prime}\oplus z_j$

\item accepts
$k_{i}=H_g(z_{1,2}^{\prime},...,z_{n,1}^{\prime},sid_i)$ as the
group session key.

\end{itemize}

\end{flushleft}

\subsubsection{P2P Stage}
\begin{flushleft}
\textbf{[P2P Key Computation]}. On input any user identity $U_j \in
pid_i$ the corresponding user $U_i$ proceeds as follows:
\begin{itemize}
\item computes $k_{i,j}^{\prime}=y_j^{x_i}=g^{x_ix_j}$,

\item accepts $k_{i,j}=H_p(k_{i,j}^{\prime},U_i|y_i,U_j|y_j)$ as the two-party session key.

\end{itemize}
\end{flushleft}

\subsection{mBD+S Protocol}
In this subsection, we briefly review the mBD+S protocol, which also
includes two stages: group stage and subgroup stage. Since the group
stage of the mBD+S protocol is same as that of the mBD+P protocol,
here we omit the details. On the correctness of key computation and
the security analysis of the mBD+S protocol refer to \cite{FGKE}.
Next, we only introduce the subgroup stage.

\subsubsection{Subgroup Stage}

On input any user identity $spid \subset pid$ the corresponding
users perform the following steps. We assume that
$spid=(U_1,...,U_m)$ with $m<n$ and that $U_0=U_m$ and
$U_{m+1}=U_1$.
\begin{flushleft}
\textbf{[Round 1]}. Each $U_i \in spid$ proceeds as follows:

\begin{itemize}
\item extracts $ssid_i=(U_1|y_1,...,U_m|y_m)$ from $sid_i$,

\item computes $k_{i-1,i}^{\prime}=y_{i-1}^{x_i}$ and
$k_{i,i+1}^{\prime}=y_{i+1}^{x_i}$,

\item $z_{i-1,i}^{\prime}=H(k_{i-1,i}^{\prime},sid_i)$ and
$z_{i,i+1}^{\prime}=H(k_{i,i+1}^{\prime},sid_i)$,

\item $z_i=z_{i-1,i}^{\prime}\oplus z_{i,i+1}^{\prime}$,

\item $\sigma_i=Sign(sk_i,(U_i,z_i,ssid_i))$,

\item broadcasts $(U_i,z_i,\sigma_i)$.

\end{itemize}

\textbf{[Subgroup Key Computation]}. Each $U_i$ checks whether
$z_1\oplus ... \oplus z_m=0$ and whether all received signatures
$\sigma_j$ are valid and aborts if any of these checks fails.
Otherwise, $U_i$ proceeds as follows:

\begin{itemize}
\item iteratively for each $j=i,...,i+m-1$, computes $z_{j,j+1}^{\prime}=z_{j-1,j}^{\prime}\oplus z_j$

\item accepts
$k_{i,J}=H_s(z_{1,2}^{\prime},...,z_{m,1}^{\prime},ssid_i)$ as the
subgroup session key.

\end{itemize}

\end{flushleft}


\section{Insider Attack on mBD+P and mBD+S Protocols}\label{IA}

In this section, we propose our attack to the group stage of their
protocols. Our attack is similar to Lee and Lee's cryptanalysis
\cite{Lee} on Jung's scheme \cite{Jung}. Under our attack, two
malicious insiders can victim a user to agree a different group
session key from other users. We note that this attack also can be
mounted to the subgroup stage in the similar way.

Suppose that users $U_{i-1}$ and $U_{i+1}$ are two malicious
insiders. They are going to deceive $U_i$ into believing that $U_i$
shares a common group session key with other users after execution
of the group stage of the mBD+P protocol or the mBD+S protocol,
while in fact $U_i$ does not have the common group session key. All
group users honestly execute the protocol during setup phase. In the
group stage, two malicious insiders $U_{i-1}$ and $U_{i+1}$ try to
disrupt the protocol as follows:

\begin{flushleft}
\textbf{[Round 1]}. Each $U_l$ (for $1\leq l\leq n$) computes
$y_l=g^{x_l}$ for some random value $x_l \in_R Z_q$ and broadcasts
$(U_l,y_l)$.

\textbf{[Round 2]}. Each $U_j$ (for $1\leq j\neq i-1,i+1\leq n$)
proceeds as follows:

\begin{itemize}
\item lets $sid_j=(U_1|y_1,...,U_n|y_n)$,

\item computes $k_{j-1,j}^{\prime}=y_{j-1}^{x_j}$ and
$k_{j,j+1}^{\prime}=y_{j+1}^{x_j}$,

\item $z_{j-1,j}^{\prime}=H(k_{j-1,j}^{\prime},sid_j)$ and
$z_{j,j+1}^{\prime}=H(k_{j,j+1}^{\prime},sid_j)$,

\item $z_j=z_{j-1,j}^{\prime}\oplus z_{j,j+1}^{\prime}$,

\item $\sigma_j=Sign(sk_j,(U_j,z_j,sid_j))$,

\item broadcasts $(U_j,z_j,\sigma_j)$ (for $1\leq j\neq i-1,i+1\leq
n$).
\end{itemize}

Malicious insider $U_{i-1}$ proceeds as follows:
\begin{itemize}
\item lets $sid_{i-1}=(U_1|y_1,...,U_n|y_n)$,

\item computes $k_{i-2,i-1}^{\prime}=y_{i-2}^{x_{i-1}}$ and
$k_{i-1,i}^{\prime}=y_{i}^{x_{i-1}}$,

\item $z_{i-2,i-1}^{\prime}=H(k_{i-2,i-1}^{\prime},sid_{i-1})$ and
$z_{i-1,i}^{\prime}=H(k_{i-1,i}^{\prime},sid_{i-1})$,

\item $z_{i-1}=z_{i-2,i-1}^{\prime}\oplus z_{i-1,i}^{\prime}\oplus r_M$,
where $r_M \in_R Z_q$ chosen by $U_{i-1}$ and $U_{i+1}$.

\item $\sigma_{i-1}=Sign(sk_{i-1},(U_{i-1},z_{i-1},sid_{i-1}))$,

\item broadcasts $(U_{i-1},z_{i-1},\sigma_{i-1})$. 

\end{itemize}

Malicious insider $U_{i+1}$ proceeds as follows:
\begin{itemize}
\item lets $sid_{i+1}=(U_1|y_1,...,U_n|y_n)$,

\item computes $k_{i,i+1}^{\prime}=y_{i}^{x_{i+1}}$ and
$k_{i+1,i+2}^{\prime}=y_{i+2}^{x_{i+1}}$,

\item $z_{i,i+1}^{\prime}=H(k_{i,i+1}^{\prime},sid_{i+1})$ and
$z_{i+1,i+2}^{\prime}=H(k_{i+1,i+2}^{\prime},sid_{i+1})$,

\item $z_{i+1}=z_{i,i+1}^{\prime}\oplus z_{i+1,i+2}^{\prime}\oplus r_M$,
where $r_M \in_R Z_q$ chosen by $U_{i-1}$ and $U_{i+1}$.

\item $\sigma_{i+1}=Sign(sk_{i+1},(U_{i+1},z_{i+1},sid_{i+1}))$

\item broadcasts $(U_{i+1},z_{i+1},\sigma_{i+1})$.
\end{itemize}

\textbf{[Group Key Computation]}. Each $U_l$ checks whether
$z_1\oplus ... \oplus z_n=0$ and whether all received signatures
$\sigma_j$ are valid and aborts if any of these checks fails.
Otherwise, all group members except victim $U_i$ proceed as follows:

\begin{itemize}
\item iteratively for each $j=l,...,l+n-1$, computes $z_{j,j+1}^{\prime}=z_{j-1,j}^{\prime}\oplus z_j$

\item accepts
$k_{l}=H_g(z_{1,2}^{\prime},...,z_{i-2,i-1}^{\prime},z_{i-1,i}^{\prime}\oplus
r_M,z_{i,i+1}^{\prime}\oplus
r_M,z_{i+1,i+2}^{\prime},...,z_{n,1}^{\prime},sid_l)$, where $1\leq
l\neq i\leq n$ as the group session key.
\end{itemize}

$U_i$ proceeds as follows:

\begin{itemize}
\item iteratively for each $j=i,...,i+n-1$, computes $z_{j,j+1}^{\prime}=z_{j-1,j}^{\prime}\oplus z_j$

\item accepts
$k_{i}=H_g(z_{1,2}^{\prime}\oplus r_M,...,z_{i-2,i-1}^{\prime}\oplus
r_M,z_{i-1,i}^{\prime},z_{i,i+1}^{\prime},z_{i+1,i+2}^{\prime}\oplus
r_M,...,z_{n,1}^{\prime}\oplus r_M,sid_i)$ as the group session key.
\end{itemize}

\end{flushleft}

Since $H_g$ is a random oracle, it is obvious that the session key
$k_{i}$ computed by $U_i$ is different from the group session key
$k_{l}$ (for $1\leq l\neq i\leq n$) computed by other users.


\section{Improvement of mBD+P and mBD+S Protocols}\label{Impro}

In this section, we propose an effective countermeasure against
malicious insider attack. The main idea to prevent the malicious
insider attack is that we add an additional round for key
confirmation to the group stage of the original mBD+P and mBD+S
protocols. In the improvement of mBD+P and mBD+S protocols
descriptions, we add two random oracles: $H_g^{'}$ is a random
oracle from $\{0,1\}^{*}$ to $\{0,1\}^{2\tau}$ and $H_{kc}$ is a
random oracle from $\{0,1\}^{*}$ to $\{0,1\}^{\tau}$. Next, we
describe the details of our improvement.

\begin{flushleft}

 \textbf{[Round 1]}. Each $U_i$ computes $y_i=g^{x_i}$ for some random
$x_i \in_R Z_q$ and broadcasts $(U_i,y_i)$.

\textbf{[Round 2]}. Each $U_i$ proceeds as follows:

\begin{itemize}
\item lets $sid_i=(U_1|y_1,...,U_n|y_n)$,

\item computes $k_{i-1,i}^{\prime}=y_{i-1}^{x_i}$ and
$k_{i,i+1}^{\prime}=y_{i+1}^{x_i}$,

\item $z_{i-1,i}^{\prime}=H(k_{i-1,i}^{\prime},sid_i)$ and
$z_{i,i+1}^{\prime}=H(k_{i,i+1}^{\prime},sid_i)$,

\item $z_i=z_{i-1,i}^{\prime}\oplus z_{i,i+1}^{\prime}$,

\item $\sigma_i=Sign(sk_i,(U_i,z_i,sid_i))$,

\item broadcasts $(U_i,z_i,\sigma_i)$.

\end{itemize}

\textbf{[Group Key Computation]}. Each $U_i$ checks whether
$z_1\oplus ... \oplus z_n=0$ and whether all received signatures
$\sigma_j$ are valid and aborts if any of these checks fails.
Otherwise, $U_i$ proceeds as follows:

\begin{itemize}
\item iteratively for each $j=i,...,i+n-1$, computes $z_{j,j+1}^{\prime}=z_{j-1,j}^{\prime}\oplus z_j$

\item computes
$(k_{i},k_{i}^{kc})=H_g^{'}(z_{1,2}^{\prime},...,z_{n,1}^{\prime},sid_i)$.
\end{itemize}

\textbf{[Key Confirmation Message]}. Each $U_i$ proceeds as follows:

\begin{itemize}
\item computes
\begin{center}
$M_i=H_{kc}(k_{i}^{kc},sid_i)$,
$\sigma_i^{kc}=Sign(sk_i,(U_i,M_i,sid_i))$ \end{center}
\item broadcasts $(U_i,M_i,\sigma_i^{kc})$.

\end{itemize}

\textbf{[Round 3]}. Each $U_i$ checks whether $M_i=M_j$ (for $1\leq
j\neq i \leq n$) and whether all received signatures $\sigma_j^{kc}$
are valid and aborts if any of these checks fails. Otherwise, $U_i$
completes the session by accepting $k_{i}$ as the common group
session key. 
\end{flushleft}

With this improvement, all group users can verify whether their
group session key are computed in the same key material and find
whether there exists malicious insiders. This simple countermeasure
is also effective to the subgroup stage of mBD+S protocol.



\section{Conclusion}\label{con}

The design of secure GKE protocols has been proved to be a
non-trivial task. Many GKE protocols had appeared in the literature
that subsequently were proved to be flawed. In this paper, we point
out that Abdalla et al.'s protocols cannot satisfy a security goal,
which is to make all group users share a common group session key.
The group stage and subgroup stage of their protocols suffer from
malicious insiders colluding attack. Two malicious insiders can
cheat a user into accepting a different session key from other
users. Further, we propose an improvement of their protocols with
key confirmation to repair this security weakness.


%


\end{document}